\documentclass[sigconf,natbib=true,anonymous=false]{acmart}

\AtBeginDocument{%
  \providecommand\BibTeX{{%
    \normalfont B\kern-0.5em{\scshape i\kern-0.25em b}\kern-0.8em\TeX}}}

\setcopyright{acmcopyright}
\copyrightyear{2018}
\acmYear{2018}
\acmDOI{XXXXXXX.XXXXXXX}

\acmConference[Conference acronym 'XX]{Make sure to enter the correct
  conference title from your rights confirmation emai}{June 03--05,
  2018}{Woodstock, NY}
\acmBooktitle{Woodstock '18: ACM Symposium on Neural Gaze Detection,
 June 03--05, 2018, Woodstock, NY} 
\acmPrice{15.00}
\acmISBN{978-1-4503-XXXX-X/18/06}

\begin{document}

\title{Rethinking Position Bias Modeling with Knowledge Distillation for CTR Prediction}

\author{Congcong Liu}
\email{liucongcong25@jd.com}
\affiliation{%
  \institution{Business Growth BU, JD.COM}
  \streetaddress{P.O. Box 1212}
  \city{Beijing}
  \country{China}
}

\author{Yuejiang Li}
\email{lyj18@mails.tsinghua.edu.cn}
\affiliation{%
  \institution{Tsinghua University}
  \city{Beijing}
  \country{China}
}

\author{Jian Zhu}
\email{zhujian146@jd.com}
\affiliation{%
  \institution{Business Growth BU, JD.COM}
  \city{Beijing}
  \country{China}
}

\author{Xiwei Zhao}
\email{zhaoxiwei@jd.com}
\affiliation{%
  \institution{Business Growth BU, JD.COM}
  \city{Beijing}
  \country{China}
}

\author{Changping Peng}
\email{pengchangping@jd.com}
\affiliation{%
 \institution{Business Growth BU, JD.COM}
  \city{Beijing}
  \country{China}
 }

\author{Zhangang Lin}
\email{linzhangang@jd.com}
\affiliation{%
  \institution{Business Growth BU, JD.COM}
  \city{Beijing}
  \country{China}
  }

\author{Jingping Shao}
\email{shaojingping@jd.com}
\affiliation{%
  \institution{Business Growth BU, JD.COM}
  \city{Beijing}
  \country{China}
  }

\renewcommand{\shortauthors}{C Liu, et al.}

\begin{abstract}
Click-through rate (CTR) Prediction is of great importance in real-world online ads systems. 
One challenge for the CTR prediction task is to capture the real interest of users from their clicked items, which is inherently biased by presented positions of items, i.e., more front positions tend to obtain higher CTR values. 
A popular line of existing works focuses on explicitly estimating position bias by result randomization which is expensive and inefficient, or by inverse propensity weighting (IPW) which relies heavily on the quality of the propensity estimation.
Another common solution is modeling position as features during offline training and simply adopting fixed value or dropout tricks when serving. However, training-inference inconsistency can lead to sub-optimal performance.
Furthermore, post-click information such as position values is informative while less exploited in CTR prediction. 
This work proposes a simple yet efficient knowledge distillation framework to alleviate the impact of position bias and leverage position information to improve CTR prediction.
We demonstrate the performance of our proposed method on a real-world production dataset and online A/B tests, achieving significant improvements over competing baseline models.
The proposed method has been deployed in the real world online ads systems, serving main traffic on one of the world's largest e-commercial platforms.

\end{abstract}




\maketitle

\section{Introduction}

Click-Through Rate (CTR) prediction is an essential task in online advertising platforms and recommender systems, which aims to correctly estimate the click probability of a user when interacting with items.
The training data of CTR models are collected from user feedback, which may be biased by the exposed position of the items.
\begin{figure}[!t]
  \includegraphics[width=0.45\textwidth]{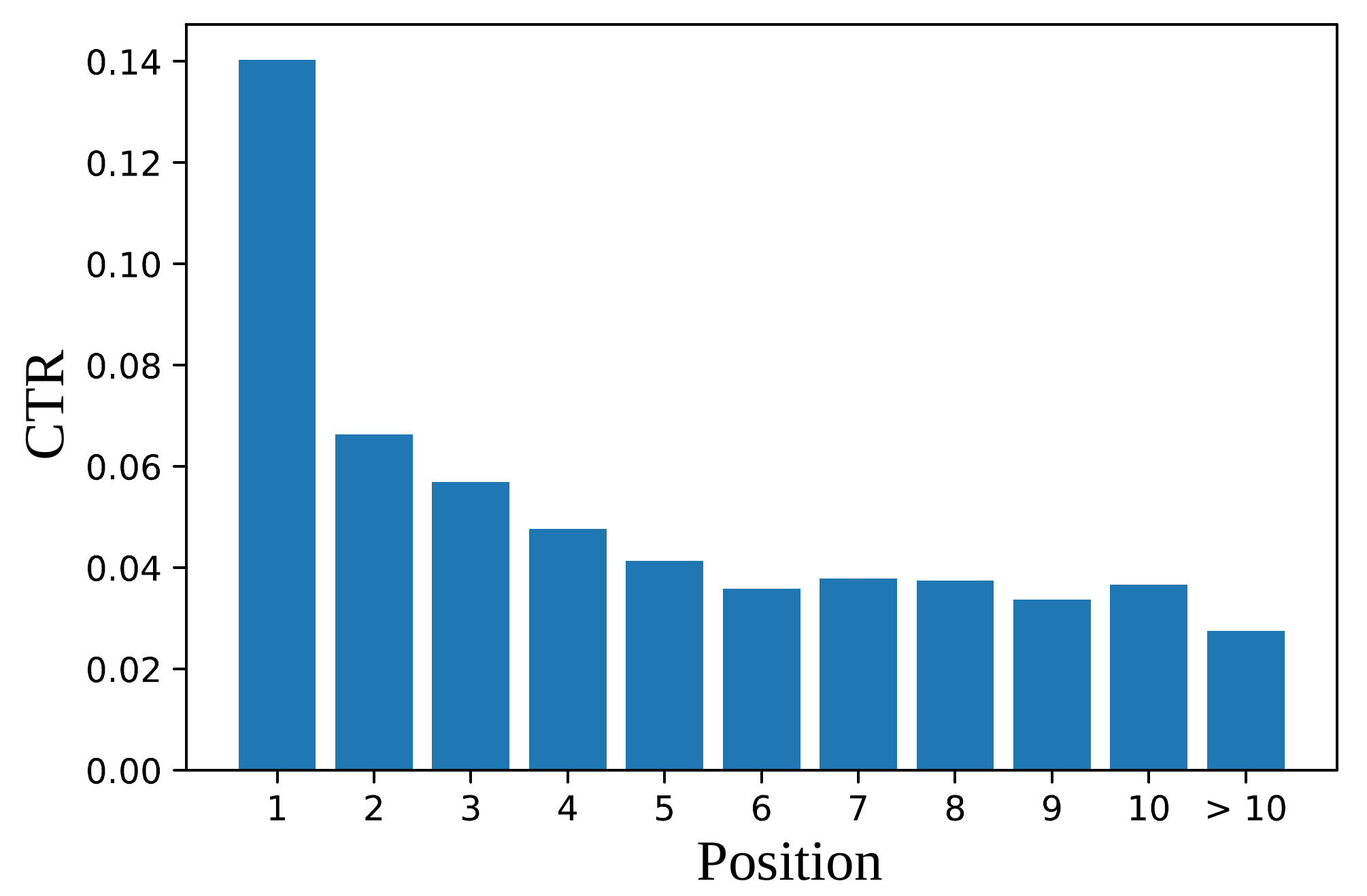}
  \caption{Real CTR distribution @ position.}
  \label{fig:ctr_gt}
\end{figure}
As shown in Figure. \ref{fig:ctr_gt}, in the real world advertising platform, we can observe the real CTR calculated by labels of click or not declines dramatically as the presented position increases.
Recent eye tracking research has shown that users tend to interact with items at higher position regardless of their relevance \cite{joachims2007evaluating,joachims2017accurately}. 
Therefore, the existence of position bias may hurt prediction accuracy and affect user's experience.
As an important factor of CTR prediction, it is important and necessary to model position bias in offline training of CTR models.

Recent years, many attempts have been made to model position bias, which closely related to performance of the CTR prediction.
A common approach is to incorporate position information as feature in the training stage, and serve with fixed position or using inverse dropout \cite{zhao2019recommending}.
However, the simple treatments for training-inference gap may lead to performance degradation. 
Another line of research tries to estimate the position bias directly by result randomization\cite{hofmann2013reusing,swaminathan2015batch,wang2016learning} or Inverse Propensity Weighting (IPW) methods \cite{ai2018unbiased,wang2018position,chen2021adapting}.
The result randomization will obviously hurt users' experience that is often not acceptable in real-world production systems, while IPW methods required the propensity scores to be properly learned.

\begin{figure*}[!t]
  \includegraphics[width=0.85\textwidth]{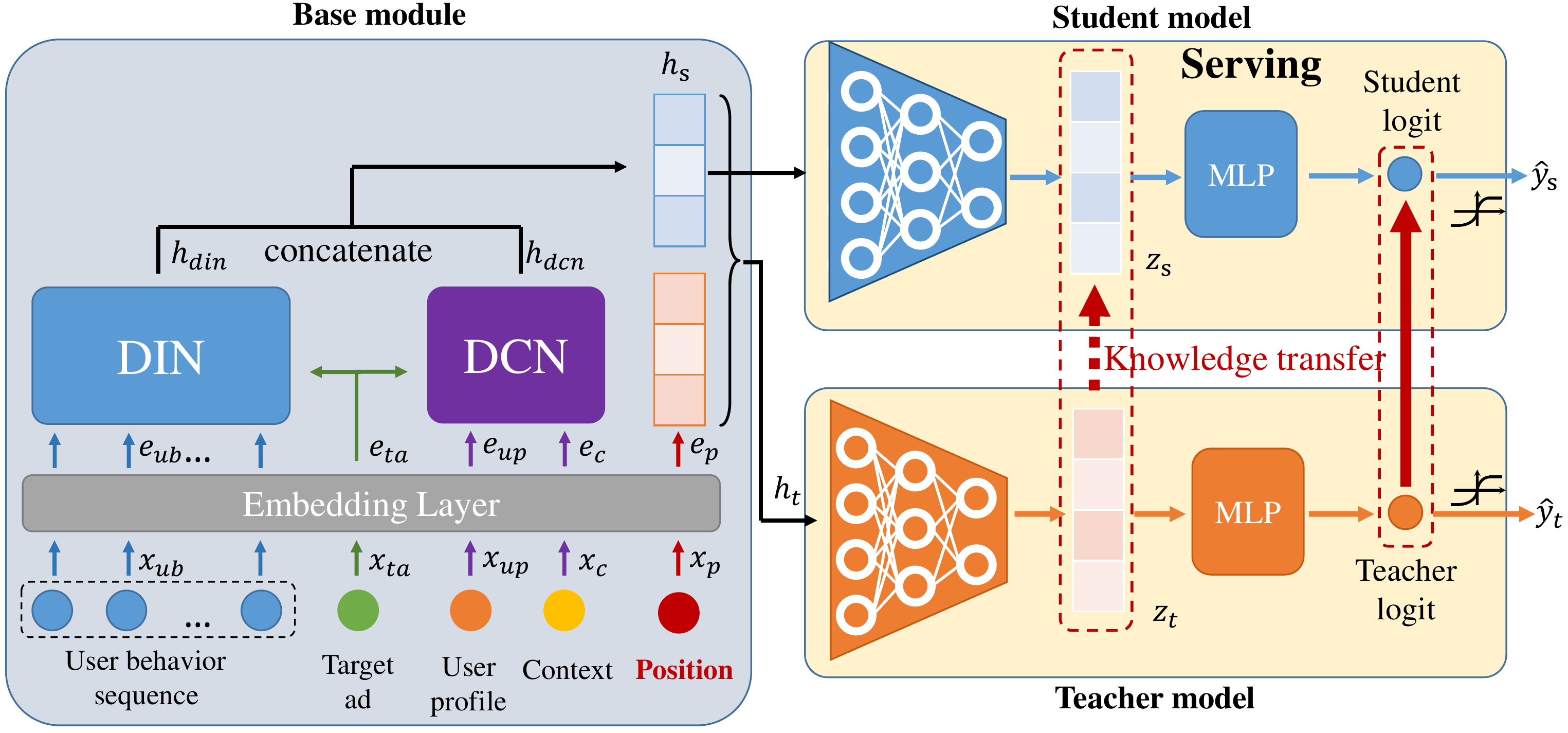}
  \caption{System diagram of the proposed knowledge distillation framework. Our framework consists of three parts, including base module, student model and teacher model. The base module takes user behavior sequence, target ad, user profile, and context info as input, and transforms them into embedding $h_s$. The Embedding layer converts the position index of items into position embedding $e_p$. When training, the $h_s$ and $e_p$ are concatenated and forwarded into teacher model, while $h_s$ is forwarded into the student model. A knowledge transfer process will be taken between the logit or features of student and teacher to guide the training process. When serving, only the student logit will be used for inference, while the position encoding part and teacher model will not participate.}
  \label{fig:system}
\end{figure*}

Furthermore, from the aspect of feature incorporation, position information contains rich information for guiding CTR prediction while has been less exploited yet.  
However, the position feature is not available during serving due to the post-click property. 
Therefore, it is important to figure out a proper way to bridge the gap between offline training and online inference of position utilization.

Knowledge distillation is a simple yet efficient way to bridge the training-inference inconsistency that has been widely applied in many areas, including machine learning, computer vision, natural language processing, etc \cite{hinton2015distilling,touvron2021training,sanh2019distilbert,jiao2019tinybert,wang2021knowledge,papernot2016distillation}.
There are two main categories of distillation forms: logit-based that focused on mimicking output logit, and feature-based that focused on exploiting intermediate layers.  
A recent work \cite{liu2020general} utilized knowledge distillation method to exploit the feature and logit from structures trained from unbiased data.
However, this method does not explicitly model the position bias, but tries to mimic the unbiased teacher models.
Also, as mentioned above, the collection of the uniform data is expensive, and obviously, the amount of data will influence the quality of the model.
In the position bias modeling of CTR prediction, knowledge distillation techniques have been less exploited.

In this paper, we propose a novel knowledge distillation framework to incorporate position information into CTR models, which alleviates the position bias and addresses the offline-online inconsistency.
We verify the superiority of our methods over various baselines on a real-world production dataset.
The rigorous online A/B tests further demonstrate the effectiveness and efficiency of our methods serving on an online advertising system.
Furthermore, an analysis of CTR distribution over positions is also provided to illustrate the debiasing effect of the proposed method.

There are several contributions of our work:
 \begin{itemize}
     \item The proposed knowledge distillation method for position bias modeling achieves significant improvements on a real-world production dataset. 
     \item We analyze the predicted logit distribution of our models and baselines over different positions to validate the effectiveness of the proposed methods to mitigate the position bias.
     \item We further examine our method with an online A/B test in a real-world ads system, achieving 1\% improvement on eCPM gain over a competitive and mature baseline model. 
     \item The proposed method has been deployed and serves main traffic on one of the world's largest e-commercial companies.
 \end{itemize}


\section{Methodology}
\label{sec:method}

In this section, we will first introduce the overview of the system diagram of this work. After that, three sub-modules will be presented in detail. Furthermore, the position bias modeling with knowledge distillation will be discussed.

\subsection{System Overview}

Figure \ref{fig:system} depicts the overall framework of the proposed method to model position information for CTR prediction.
There are three parts in our framework: base module, student model, teacher model.
The base module takes various task-related features as input, including user behavior sequence, target ad, user profile, etc. These information will be transformed into embeddings for further process.
The teacher model will participate in the training process to guide the student model for better CTR prediction. When serving, only the student model will be used for inference.

\subsection{Base Module}
As shown in the left part of Figure \ref{fig:system}, the base module consists of two main parts: the DIN module for user behavior interaction, the DCN module for feature interaction.
DIN module takes user behavior sequences $x_{ub}$ and target ad $x_{ta}$ as input to extract the user behavior information $h_{din}$.
\begin{equation}
    h_{din} = DIN(x_{ub},x_{ta})
\end{equation}
The DCN module converts user profile $x_{up}$, context information $x_c$ and also target ad $x_{ta}$ into feature interaction embedding $h_{dcn}$ \begin{equation}
    h_{dcn} = DIN(x_{up},x_{c},x_{ta})
\end{equation}
The embeddings generated from DIN and DCN will be concatenated for further process.
\begin{equation}
    h_{s} = Concatenate(h_{din},h_{dcn})
\end{equation}

The Embedding Layer converts the position index of items into position embedding $e_p$ when training.

\subsection{Teacher Model}
The teacher model takes the overall feature embedding $h_{s}$ and position embedding $e_p$ as input. A DNN based encoder transforms the input into feature embedding $z_t$.
\begin{equation}
    z_t = Encoder(Concatenate(h_s,e_p))
\end{equation}
After that, the $z_t$ will be fed into MLP with a sigmoid activation function to obtain the predicted CTR $y_t$.
\begin{equation}
    \hat{y}_t = \sigma(MLP(z_t))
\end{equation}

\subsection{Student Model}
Different from the teacher model, only the overall feature embedding $h_{s}$ will be forwarded into the student model.
Similarly, a DNN based encoder will convert the $h_s$ into embedding $z_s$.
\begin{equation}
    z_s = Encoder(Concatenate(h_s))
\end{equation}
Then the $z_s$ will be forwarded into MLP with a sigmoid activation function to obtain the predicted CTR $y_s$ of student model.
\begin{equation}
    \hat{y}_s = \sigma(MLP(z_s))
\end{equation}
\subsection{Position Bias Modeling via Knowledge Distillation}
\subsubsection{Training}
As shown in the right part of the Figure \ref{fig:system}, the overall feature embedding $h_{s}$ and position embedding $e_p$ are concatenated and forwarded into the teacher model. $h_{s}$ is fed into the student model.

During training, given that the predicted CTR of teacher model is $\hat{y}_t$ and the ground truth label is $y_g\in\{0, 1\}$, we adopt the cross entropy loss \begin{equation}
    CE(\hat{y}_t, y_g) = y_g \log(\hat{y}_t) + (1 - y_g) \log(1 - \hat{y}_t),
\end{equation}
to train teacher model.
For the student model, it is trained from two perspectives.
First, similar to teacher model, we also adopt the cross-entropy loss to student model with the ground truth label, that is,
\begin{equation}
    CE(\hat{y}_s, y_g) = y_g \log(\hat{y}_s) + (1 - y_g) \log(1 - \hat{y}_s).
\end{equation}
We also defined the above two cross entropy loss as point-wise loss.

Apart from the above cross-entropy loss for the student model, since the teacher model incorporates the position information, the output of intermediate layer and logit of teacher model contain more information than those of student model.
To impose that student model can learn such additional information of position feature from the teacher model, we adopt the knowledge distillation methods to achieve this. Specifically, we adopt two distillation methods for student model: \textit{logit-based} distillation and \textit{feature-based} distillation, respectively.
For logit-based distillation, we aim to make the output of student model to be similar to that of teacher model.
Thus, we use the predicted CTR of teacher model, $\hat{y}_t$, as the soft label, and add the logit distillation loss
\begin{equation}
    CE(\hat{y}_s, \hat{y}_t) = \hat{y}_t \log(\hat{y}_s) + (1 - \hat{y}_t) \log(1 - \hat{y}_s),
\end{equation}
for student model. While for feature-based distillation, we aim to make the feature embedding of teacher model and that of student model to be as similar as possible.
To this end, we adopt the mean squared error between $z_t$ and $z_s$ for feature-based distillation, that is,
\begin{equation}
    MSE(z_s, z_t) = \Vert z_s - z_t \Vert_2^2.
\end{equation}

Consequently, we can summarize the whole loss of our architecture as
\begin{equation}
    loss = CE(\hat{y}_s, y_g) + CE(\hat{y}_t, y_g) + \lambda \cdot distill\_loss,
\end{equation}
where $distill\_loss$ can be $CE(\hat{y}_s, \hat{y}_t)$ with logit-based distillation or $MSE(z_s, z_t)$ with feature-based distillation. $\lambda$ denotes the hyperparameter that controls the effect of knowledge distillation. 


In real deployment, logit-based method is selected with better performance which will be discussed in Section. \ref{sec:exp}.




Through the knowledge distillation process, the position information inside the teacher model will be transferred into the student model for serving.
\subsubsection{Serving}
Only the overall feature embedding $h_s$ that forwarded into student model will be used when serving.
The position embedding $e_p$ and the teacher model will not participate in the inference stage.

\section{Experiment}
\label{sec:exp}
In this section, we describe how we conduct our experiments of the proposed knowledge distillation framework for position bias modeling in CTR prediction task.
Using the user feedback collected from a real-world online advertising platform, we train the CTR models and conduct offline experiments against various baselines.
An online A/B test is also deployed in the ranking system of the online advertising platform.

\subsection{Experimental Settings}
\subsubsection{Datasets}

A real-world production dataset collected from the user feedback data on an online advertising system in one of the world's largest e-commercial companies is adopted for experiments.
The real-world production dataset contains about ten billion records collected in one month (day 1 - day 30).
A validation set is constructed by extracting about 0.5 million records from the ten billion records. The rest is used as the training set.
For testing, about one million records collected from the following day (day 31) are used for evaluation of the proposed method and baselines.

\subsubsection{Evaluation Metrics}
Similar to prior works, we adopt AUC (Area Under ROC) and LogLoss as the evaluation metrics in offline experiments.
Higher AUC and lower Logloss denote better performance of the models.
The result reported in this work is averaged over 5-runs.

\subsection{Baselines}
To evaluate the proposed knowledge distillation method for position bias modeling, we compare it with the following baseline methods.
The baseline models are developed upon ensembles of a popular CTR model DIN \cite{zhou2018deep} and DCN \cite{wang2017deep} as the backbone.

\begin{itemize}
    \item \textbf{Backbone}. An ensemble of DIN and DCN is used as backbone.
    \item \textbf{Backbone + Fixed Position}. This method takes the position as input features in training and use the first position for serving.
    \item \textbf{Backbone + PosDropOut \cite{zhao2019recommending}}. This method takes the position as a feature and applies dropout trick. Dropout rate here is 0.1, consistent with \cite{zhao2019recommending}.
    \item \textbf{Backbone + PAL \cite{guo2019pal}}.  PAL method is applied upon the backbone structure.
\end{itemize}

\subsection{Offline experiment}

\subsubsection{Comparing to baseline methods}

\begin{table}[]
\centering
\caption{ Offline results in real-world industrial datasets over 5-runs results. Std$\approx$1e-3. }
\label{tab:offline_benchmark}
\begin{tabular}{lcc}
\hline
Model                & AUC     & LogLoss  \\ \hline
Backbone                  &  0.7473    & 0.5092     \\
Backbone + Fixed Position & 0.7477   & 0.5087    \\
Backbone + PosDropOut     & 0.7483    & 0.5082     \\
Backbone + PAL            & 0.7485    & 0.5080     \\
Ours                 & 0.7528    & 0.5050     \\ \hline
\end{tabular}
\end{table}

Table \ref{tab:offline_benchmark} shows the comparisons of the proposed method with various baselines.
From the table, we can see that the proposed method overwhelm all baseline methods significantly with the highest AUC and lowest Logloss.

When compared to the Backbone method with an ensemble of DCN and DIN which is highly optimized with large number of useful features and attributes, our method achieves 0.74\% improvement on AUC and 0.82\% improvement on Logloss.

When comparing to other position bias modelling methods, our method achieves 0.57\% - 0.68\% improvements on AUC and 0.59\% - 0.72\% improvements on LogLoss.

\subsubsection{Logit-based distillation vs. feature-based distillation}
\begin{table}[]
\centering
\caption{Ablation study on different knowledge distillation methods. Std$\approx$1e-3. }
\label{tab:kd_ablation}
\begin{tabular}{lcc}
\hline
Model               & AUC       & LogLoss \\ \hline
Logit-based         & 0.7528      & 0.5050    \\
Feature-based       & 0.7484      & 0.5081    \\ \hline
\end{tabular}
\end{table}

We conduct experiments to investigate two common forms of knowledge distillation methods: logit-based (or response-based) that exploits the output logit, and feature-based method \cite{gou2021knowledge} that exploits the intermediate features.
We sweep the hyper-parameter $\lambda$ from \{$1e^{-2}$, $1e^{-1}$, $2e^{-1}$, $4e^{-1}$, $6e^{-1}$, $8e^{-1}$, 1\}.
We set the $\lambda$ which yields the lowest loss in the validation set, which is 1 for logit-based KD and $6e^{-1}$ for feature-based KD.  
Table. \ref{tab:kd_ablation} illustrate the performance of logit-based and feature-based knowledge distillation.
It is clear to see that the logit-based method significantly outperforms the feature-based method with a large margin of 0.58\% improvements in AUC and 0.61\% decrease in LogLoss. 
This observation is interestingly consistent with \cite{duong2019shrinkteanet,romero2014fitnets} due to potential over-regularized issue.

\subsection{Online A/B tests}

\begin{table}[]
\centering
\caption{Results of Online A/B testing.}
\label{tab:onlineexp}
\begin{tabular}{lccc}
\hline
Online   & CTR Gain & eCPM Gain & TP99 latency \\ \hline
Base model   & 0\%     & 0\%     & 25ms      \\ 
Ours     & 1.0\% $\uparrow$   & 1.07\% $\uparrow$  & 25ms \\ \hline
\end{tabular}
\end{table}

From 2022-Feb-12 to 2022-Feb-18, the online A/B testing was conducted on the ranking system of a real-world online advertising platform.
From Table. \ref{tab:onlineexp}, the proposed method achieves 1.0\% CTR gain and 1.07\% eCPM (Effective Cost Per Mille) gain compared to the previous online base model, without bringing extra latency.
\subsection{Qualitative result}

\begin{figure}[!t]
  \includegraphics[width=0.40\textwidth]{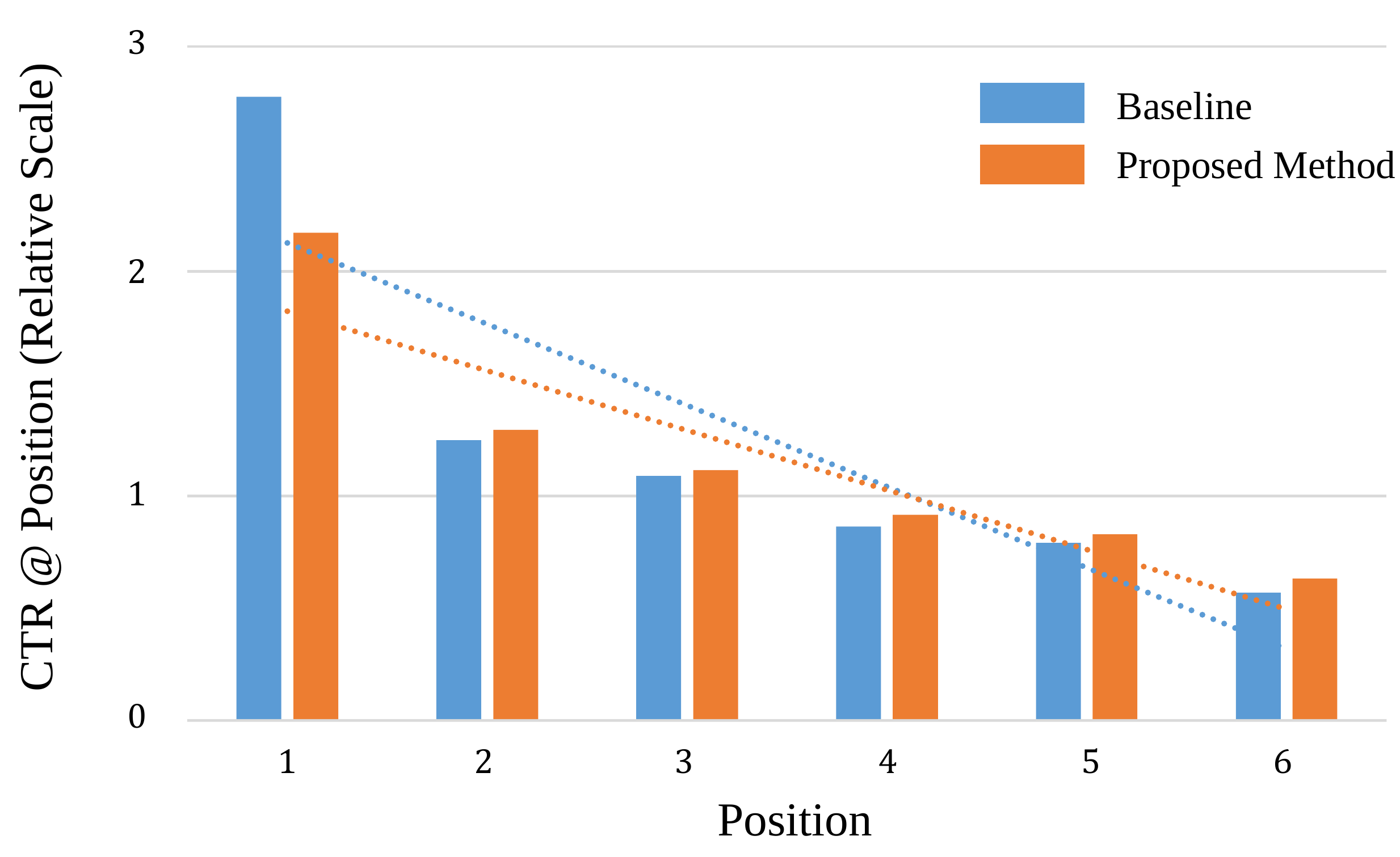}
  \caption{pCTR @ pos}
  \label{fig:ctr}
\end{figure}

Figure \ref{fig:ctr} illustrates the pCTR distribution over different positions between the proposed method and baseline method.
As expected, we can observe the decaying effect of the pCTR distribution when the position gets lower, which is similar to ground truth CTR in Figure \ref{fig:ctr_gt}.
Higher positions have higher CTR values due to a compound effect of relevance and position bias.
It is interesting to find that the pCTR values generated from the proposed method are lower at the first position while higher at other positions. 
This finding further demonstrates the effectiveness of the proposed method for position bias modeling.

\section{Conclusion}
\label{sec:conclusion}
In this work, we propose an innovative knowledge distillation framework for position bias modeling in CTR prediction.
The offline experiments on an industrial production dataset and online A/B test on an online advertising system show the superiority and efficiency of the proposed method against various baseline models for CTR prediction.
The proposed knowledge distillation framework outperforms all baseline models with position bias modeling by a significant margin on the metrics of AUC and LogLoss.
An analysis of the pCTR distribution of the proposed method and baseline method further demonstrates the effectiveness of position bias modeling with the knowledge distillation framework.
Different distillation forms, including logit-based and feature-based are also studied.
We find that the logit-based knowledge distillation method achieves better performance on CTR prediction since the feature-based method might suffer from over-regularization problem.


\bibliographystyle{ACM-Reference-Format}
\bibliography{sample-base}










\end{document}